\documentclass[aps,prx,superscriptaddress,reprint]{revtex4-1}
\usepackage{amsmath}
\usepackage{amssymb}
\usepackage{graphicx}
\usepackage{url}
\usepackage{hyperref}
\usepackage{color,xcolor}
\usepackage{epstopdf}
\usepackage{float}
\usepackage{ulem}
\usepackage[percent]{overpic}
\usepackage{siunitx}

\begin{document}

\title{Prediction of orbital selective Mott phases and block magnetic states \\ in the
quasi-one-dimensional iron chain Ce$_2$O$_2$FeSe$_2$ under hole and electron doping}
\author{Ling-Fang Lin}
\author{Yang Zhang}
\affiliation{Department of Physics and Astronomy, University of Tennessee, Knoxville, Tennessee 37996, USA}
\author{Gonzalo Alvarez}
\affiliation{Computational Sciences \& Engineering Division and Center for Nanophase Materials Sciences, Oak Ridge National Laboratory, Oak Ridge, TN 37831, USA}
\author{Jacek Herbrych}
\affiliation{Department of Theoretical Physics, Faculty of Fundamental Problems of Technology,
Wroc\l{}aw University of Science and Technology, 50-370 Wroc\l{}aw, Poland}
\author{Adriana Moreo}
\author{Elbio Dagotto}
\affiliation{Department of Physics and Astronomy, University of Tennessee, Knoxville, Tennessee 37996, USA}
\affiliation{Materials Science and Technology Division, Oak Ridge National Laboratory, Oak Ridge, Tennessee 37831, USA}

\begin{abstract}
The recent detailed study of quasi-one-dimensional iron-based ladders, with the $3d$ iron electronic density $n = 6$, has unveiled surprises, such as orbital-selective phases. However, similar studies for $n=6$ iron chains are still rare. Here, a three-orbital electronic Hubbard model was constructed to study the magnetic and electronic properties of the quasi-one-dimensional $n=6$ iron chain Ce$_2$O$_2$FeSe$_2$, with focus on the effect of doping. Specifically, introducing the Hubbard $U$ and Hund $J_{H}$ couplings and studying the model via the density matrix renormalization group, we report the ground-state phase diagram varying the electronic density away from $n=6$. For the realistic Hund coupling $J_{H}/U = 1/4$, several electronic phases were obtained, including a metal, orbital-selective Mott, and Mott insulating phases.
Doping away from the parent phase, the competition of many tendencies leads to a variety of magnetic states, such as ferromagnetism, as well as several antiferromagnetic and magnetic ``block" phases. In the hole-doping region, two different interesting orbital-selective Mott phases were found: OSMP1 (with one localized orbital and two itinerant orbitals) and OSMP2 (with two localized orbitals and one itinerant orbital). Moreover, charge disproportionation phenomena were found in special doping regions. We argue that our
predictions can be tested by simple modifications in the original chemical formula of Ce$_2$O$_2$FeSe$_2$.
\end{abstract}

\maketitle

\section{Introduction}
Since the discovery of superconductivity in iron pnictides LaFeAsO~\cite{Kamihara:Jacs}, iron-based compounds with the Fe$X_4$ tetrahedra structure ($X$ = pnictides or chalcogens) rapidly developed into one of the main branches of unconventional superconductivity~\cite{stewart2011superconductivity,Dai:Np,Dagotto:Rmp,Dai:Rmp,hosono2018recent}. In contrast to having only one active orbital as in copper-based superconductors~\cite{Dagotto:Rmp94}, the iron-based superconductors require a multi-orbital description~\cite{Scalapino:rmp}.

Different from the canonical N\'eel antiferromagnetic (AFM) order in planar copper-based superconductors involving staggered spins in both directions, the magnetism of the nonsuperconducting parent state of the quasi-two-dimensional (2D) iron-based superconductors can involve exotic magnetic phases~\cite{Dai:Np}
because of its multi-orbital nature. These many magnetic states include the collinear stripe-like AFM order (C-type AFM), which is dominant in most iron pnictides with the FeAs-layered structure~\cite{Cruz:Nat,huang2008neutron,li2009structural}. However, other states were unveiled, such as the bi-collinear AFM order in FeTe~\cite{bao2009tunable,Li:Prb09}, and the block AFM order in $A_y$Fe$_{1.6+x}$Se$_2$ with regularly spaced iron vacancies~\cite{Bao:Cpl,Ye:Prl}. Furthermore, many striking phenomena were reported in iron superconductors, primarily induced by its orbital selective character~\cite{yin2011kinetic}. For example, orbital-dependent band renormalizations~\cite{liu2015experimental,yi2015observation}, orbital-selective quasiparticles~\cite{kostin2018imaging}, orbital-selective Cooper pairing~\cite{sprau2017discovery}, and others~\cite{caron2012orbital,li2017competing} were reported in experimental studies of iron superconductors.

Another exotic example of orbital sensitive characteristics is represented by the orbital selective Mott phase (OSMP)~\cite{de2009orbital}, which could play an important
role to understand pairing in iron-based superconductors~\cite{yin2011kinetic,Yi:prl13,zhang2012general,yu2013orbital,de2014selective,yu2020orbital,yi2015observation,caron2012orbital}. These orbital selective characteristics were also proposed for the recently discovery 2D nickelate superconductor~\cite{li2019superconductivity}, where the {$d_{z^2}$ orbital is itinerant while the $d_{x^2-y^2}$ orbital displays Mott behavior~\cite{zhang2020similarities,Lechermann:prx}.

Similarly as in one-dimensional (1D) copper ladders, superconductivity under pressure was also observed in the two-leg quasi-1D iron-based ladder system BaFe$_2$X$_3$ ($X$ = S, Se) with electronic density $n = 6.0$~\cite{Takahashi:Nm,Ying:prb17}. Under ambient conditions, BaFe$_2$S$_3$ displays a stripe-type AFM order below $120$~K, similar to the C-type AFM in iron layer superconductors, involving AFM legs and ferromagnetic (FM) rungs, effectively forming $2\times1$ blocks~\cite{Takahashi:Nm}. Replacing S by Se, BaFe$_2$Se$_3$ displays an exotic AFM state with $2\times2$ FM blocks coupled antiferromagnetically along the long ladder direction below $256$~K ~\cite{caron2012orbital}. Furthermore, the OSMP was argued to be relevant for such compounds~\cite{Craco:prb20,Mourigal:prl}. Both these block OSMP states were theoretically predicted before experiments confirmed their existence. Specifically, ladder and chain iron-based materials have recently been systematically simulated
computationally~\cite{herbrych2018spin,patel2019fingerprints,herbrych2019novel,patel2020emergence,pandey2021intertwined,sroda2021quantum}, using multiorbital Hubbard models and the density matrix renormalization group (DMRG) algorithm,
unveiling a variety of new states. These interesting developments in the area of two-leg iron ladder systems, involving both theory and experiments, naturally introduce a simple question: do iron {\it chains} with $n = 6$ display similarly interesting physical properties? To address theoretically this issue,
a specific $n=6$ chain must be chosen, as described below.

\begin{figure}
\centering
\includegraphics[width=0.48\textwidth]{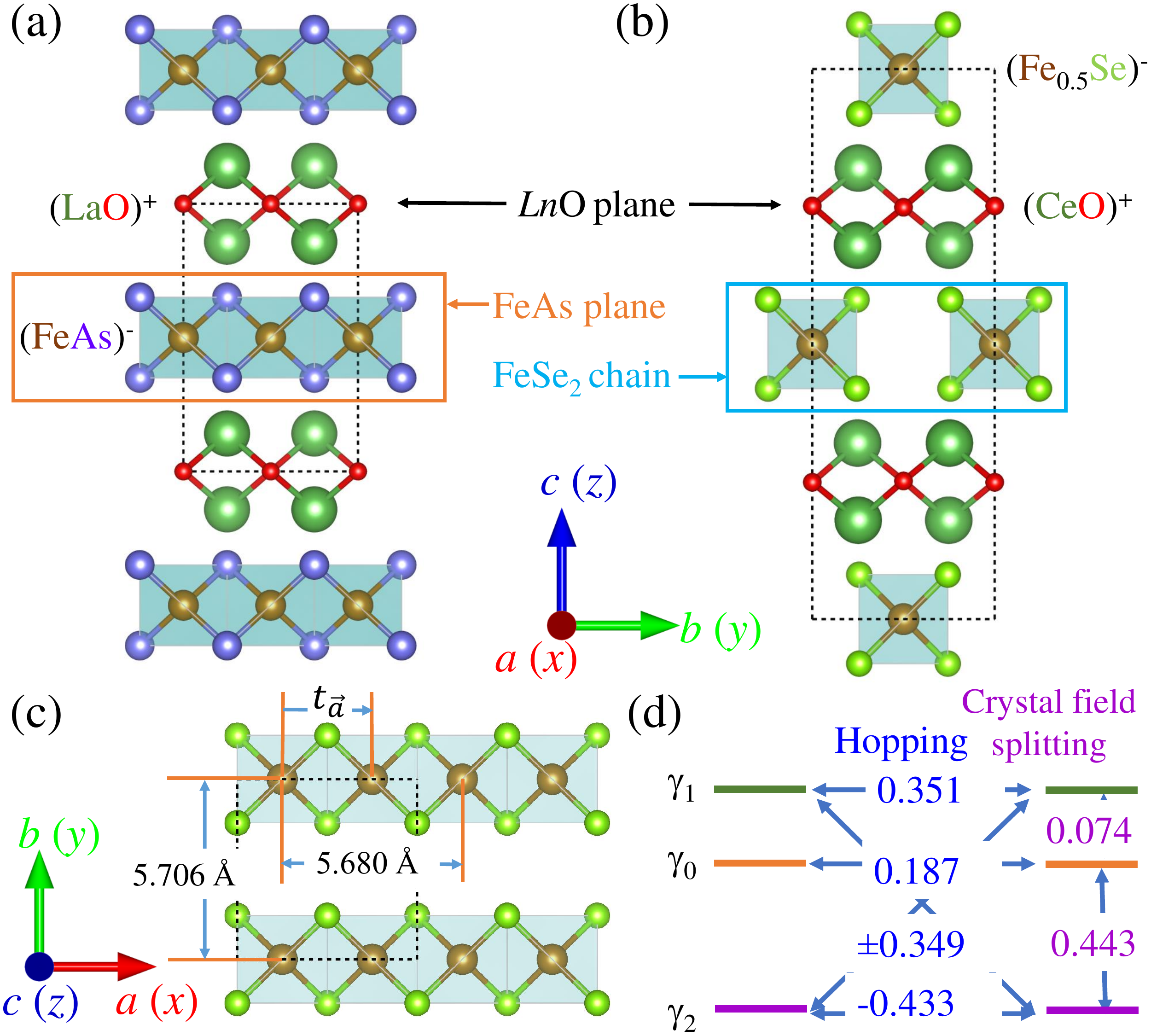}
\caption{Comparison of the crystal structures of (a) the canonical 2D superconductor parent compound $Ln$FeAsO and (b) the quasi-1D $Ln_2$O$_2$FeSe$_2$ compound of our focus. (c) Top view of the FeSe$_2$ chains present in $Ln_2$O$_2$FeSe$_2$. Here COFS is the chosen example to display specific interatomic distances and hopping amplitudes. Note that the hopping matrix along the opposite direction as shown should be the transpose one $t^{\rm T}_{\vec{a}}$. (d) Sketches of the calculated crystal-field splitting and the dominant hopping parameters for the three-orbitals \{$d_{z^2}$, $d_{yz}$, $d_{x^2-y^2}$\}, denoted as \{$\gamma_0$, $\gamma_1$, $\gamma_2$\} for simplicity.}
\label{structure}
\end{figure}

Different from the previously well-studied iron ladders~\cite{Zhang:prb17,Zhang:prb18,Zheng:prb18,Wu:prb19,Craco:prb20,zhang2019magnetic,herbrych2020block,herbrych2020block1,zhang2020iron,herbrych2021interaction}, $n=6$ iron chains have been rarely studied, in both theory and experiments. To our best knowledge, there have been only a few iron chalcogenide chains with $n = 6$ experimentally prepared or theoretically studied, such as Ba$_2$FeS$_3$ ~\cite{duan2021high,zhang2021magnetic}, Na$_2$Fe$X_2$ ($X$ = S, Se)~\cite{stuble2018na7,pandey2020prediction} and $oI$-$Ln_2$O$_2$FeSe$_2$ ($Ln$ = Ce, La) \cite{mccabe2011new,mccabe2014magnetism,stock2016magnetic,nitsche2014new}.

Interestingly, the family of materials $oI$-$Ln_2$O$_2$FeSe$_2$ (SG: Imcb) (see Fig.~\ref{structure}), which is structurally related to the iron-pnictide superconductor family $Ln$FeAsO, have been
more extensively investigated experimentally than theoretically~\cite{mccabe2011new,mccabe2014magnetism,stock2016magnetic,nitsche2014new}. Here, Ce$_2$O$_2$FeSe$_2$ (COFS) is used  as example for a detailed theoretical analysis. Replacing arsenic by selenium leads to an atomic structure where only half the iron positions are occupied to maintain charge compensation,
forming an array of iron chains instead of a plane, as shown in Figs.~\ref{structure}(a) and (b). Each chain is made of edge-sharing FeSe$_4$ tetrahedra. Experiments indicate that COFS is a promising quasi-1D chain system~\cite{mccabe2014magnetism} due to the Fe-Se-Ce interactions being much weaker than the Fe-Se-Fe nearest neighbor (NN) interactions along the dominant chain. In this chain direction, the spins couple in a FM arrangement with a large magnetic moment $\sim 3.3~\mu_{\rm B}$~\cite{mccabe2014magnetism}. In addition, COFS is an insulator with a band gap $\sim 0.64$ eV, located at the insulating side of the Mott boundary~\cite{mccabe2011new,mccabe2014magnetism}. Based on an intuitive second-order perturbation theory analysis and detailed DMRG calculations, a previous study revealed that large entanglements between doubly occupied and half-filled orbitals play a key role in stabilizing FM order along the chain direction for COFS~\cite{lin2021origin}. This provided a novel mechanism to induce FM order.

Considering related developments of orbital-selective magnetism and OSMP under carrier doping in iron ladders, such as (K, Ba)Fe$_2$Se$_3$~\cite{caron2012orbital}, the next natural step for iron chains is to understand the magnetism and OSMP varying the electronic density i.e. doping the parent compound. To address this important aspect, in this work the DMRG method was employed to investigate the magnetism and underlying electronic state properties based on the real quasi-1D $n=6$ COFS material. Based on our previous Wannier functions obtained from first principles calculations~\cite{lin2021origin}, we present a multi-orbital Hubbard model for the iron chains.

Next, we calculate the ground-state phase diagram using DMRG varying the on-site Hubbard repulsion $U$ and the carrier density, with the realistic on-site Hund coupling $J_H/U = 1/4$.  A variety of electronic states were revealed, including a canonical metal (M), an orbital selective Mott phase (OSMP), and a Mott insulator (MI). The latter becomes stable at large $U/W$ for {\it integer} electronic numbers. Moreover, rich magnetic states were also obtained in our DMRG phase diagram, involving FM, several different AFM's, as well as ``block" patterns, similar to those predicted in a related context but using hopping amplitudes resembling the planar iron superconductors~\cite{herbrych2018spin,patel2019fingerprints,herbrych2019novel}.
Interestingly, due to the strong {\it interorbital} hoppings between double and single occupied orbitals, ferromagnetism dominates most density regions at large Hubbard $U$, either for hole or electron doping away from $n=4$. Furthermore, in the regime of intermediate Hubbard coupling strengths, and within the hole dopping regime, OSMP physical properties were found, accompanied by several novel magnetic states, including block states. Within OSMP, the competition between FM and AFM exchange is the key to stabilize those block states, favoring antiferromagnetically coupled ferromagnetic islands~\cite{herbrych2019novel}.

\section{Multiorbital Hubbard model}

Several exotic phenomena have been unveiled theoretically in low dimension, including novel magnetic states, orbital ordering, ferroelectricity, nodes in the spin density, as well as dimerization~\cite{gao2020weakly,pandey2021origin,lin2021orbital,lin2019frustrated,lin2019quasi,lin2021oxygen,zhang2021peierls,zhang2021orbital,zhang2020first}. These interesting phenomena can all be qualitatively described using strictly 1D models. The novel effects mentioned above are in part due to enhanced quantum fluctuations found within the 1D systems.

Hence, to better understand the magnetic coupling of the COFS material under hole and electron doping away from
the reference density $n=4$, here an effective three-orbital Hubbard model for a quasi-1D Fe chain model was constructed. The electronic density must represent the realistic case where the valence of iron is Fe$^{\rm 2+}$ ($n$ = 6)~\cite{rincon2014exotic,rincon2014quantum,daghofer2010three}. This leads to four electrons in the active three orbitals, approximation shown to provide an excellent description of the physical properties of real iron systems with $n= 6$~\cite{rincon2014exotic,rincon2014quantum,daghofer2010three}.

Specifically, the kinetic energy and interaction energy terms $H = H_k + H_{int}$ are both included in the model. The tight-binding kinetic component is
\begin{eqnarray}
H_k = \sum_{\substack{i\sigma\\{a}\gamma\gamma'}}t_{\gamma\gamma'}^{{a}}
(c^{\dagger}_{i\sigma\gamma}c^{\phantom\dagger}_{i+{a}\sigma\gamma'}+H.c.)+ \sum_{i\gamma\sigma} \Delta_{\gamma} n_{i\gamma\sigma},
\end{eqnarray}
where the first term represents the hopping of an electron from orbital $\gamma$ at site $i$ to orbital $\gamma'$ at the NN site $i+{a}$, while $\gamma$ and $\gamma'$ represent the three different orbitals. For simplicity, only the most important NN hopping amplitudes are included in our model and the hopping matrix we used is given by (eV units)
\begin{equation}
\begin{split}
t_{{a}} =
\begin{bmatrix}
          0.187	    &  -0.054	   &       0.020	   	       \\
          0.054	    &   0.351	   &      -0.349	   	       \\
          0.020	    &   0.349	   &      -0.433	
\end{bmatrix}.\\
\end{split}
\end{equation}
$\Delta_{\gamma}$ is the crystal-field splitting of orbital $\gamma$, i.e., $\Delta_{0} = -0.277$, $\Delta_{1} = -0.203$, and $\Delta_{2} = -0.720$ eV. See Figs.~\ref{structure}(c) and (d) where the chain structure and the hoppings are displayed. The total kinetic energy bandwidth is $W=2.085$~eV. All parameters mentioned above, namely the hopping matrix and crystal-field splitting, are extracted from first-principles density functional theory (DFT) calculations~\cite{Kresse:Prb99,Blochl:Prb2,Perdew:Prl08}, supplemented by maximally localized Wannier functions~\cite{marzari1997maximally,mostofi2008wannier90}. The readers are referred to our previous work for additional details~\cite{lin2021origin}.

The electronic interaction portion of the Hamiltonian, including the standard intraorbital Hubbard repulsion, the electronic repulsion between electrons at different orbitals, Hund's coupling, and pair hopping terms, is written as:
\begin{eqnarray}
H_{int}= U\sum_{i\gamma}n_{i\uparrow \gamma} n_{i\downarrow \gamma} +(U'-\frac{J_H}{2})\sum_{\substack{i\\\gamma < \gamma'}} n_{i \gamma} n_{i\gamma'} \nonumber \\
-2J_H  \sum_{\substack{i\\\gamma < \gamma'}} {{\bf S}_{i,\gamma}}\cdot{{\bf S}_{i,\gamma'}}+J_H  \sum_{\substack{i\\\gamma < \gamma'}} (P^{\dagger}_{i\gamma} P_{i\gamma'}+H.c.).
\end{eqnarray}
where the standard relation $U'=U-2J_H$ is assumed and $P_{i\gamma}$=$c_{i \downarrow \gamma} c_{i \uparrow \gamma}$.

To investigate the properties of the COFS quasi-1D system, DMRG methods~\cite{white1992density,white1993density,schollwock2005density,hallberg2006new} were employed to address the Hamiltonian numerically using the DMRG++ computer program~\cite{alvarez2009density}.
In this DMRG calculations, we used an $L = 16$ sites cluster chain with open-boundary conditions (OBC). Furthermore, at least $m=1200$ states were employed and up to 21 sweeps were performed during this finite-size algorithm evolution. Truncation error remained below $10^{-6}$ for all of our results.

To identify different phases, several expectation values and two-point correlation functions were calculated. For example, the site-average occupation number of each orbital is defined as:
\begin{eqnarray}
n_{\gamma} =\frac{1}{L}\sum_{\substack{i,\sigma}} \langle n_{i\sigma \gamma}\rangle.
\end{eqnarray}
The orbital-resolved charge fluctuation is:
\begin{eqnarray}
\delta n_{\gamma} =\frac{1}{L}\sum_{\substack{i}}(\langle n_{i,\gamma}^2\rangle-\langle n_{i,\gamma}\rangle^2).
\end{eqnarray}
The mean value of the squared spin for each orbital is:
\begin{eqnarray}
\langle {\bf{S}}^2\rangle_{\gamma} =\frac{1}{L}\sum_{\substack{i}} \langle {\bf{S}}_{i,\gamma}\cdot{\bf{S}}_{i,\gamma}\rangle.
\end{eqnarray}
The spin correlation is defined as
\begin{eqnarray}
S_{i,j}=\langle {\bf{S}}_{i}\cdot {\bf{S}}_{j}\rangle,
\end{eqnarray}
where ${\bf{S}}_{i} =\sum_{\substack{\gamma}}{\bf{S}}_{i,\gamma}$.
The charge correlation is defined as
\begin{eqnarray}
n_{i,j}=\langle n_{i}n_{j}\rangle.
\end{eqnarray}
where $n_{i} =\sum_{\substack{\gamma}}n_{i,\gamma}$.
Finally, the corresponding structure factors for spin and charge are
\begin{eqnarray}
S(q)=\frac{1}{L}\sum_{\substack{j,m}}e^{-iq(j-m)}\langle {\bf{S}}_{m}\cdot {\bf{S}}_{j}\rangle,
\end{eqnarray}
\begin{eqnarray}
N(q)=\frac{1}{L}\sum_{\substack{j,m}}e^{-iq(j-m)}\langle (n_{m}-n)\cdot (n_{j}-n)\rangle.
\end{eqnarray}
where $n$ is the electronic density we are investigating.

\section{Results}

\subsection{Phase diagram under doping}
Before describing the DMRG results modifying the carrier doping, the basic results corresponding to COFS without doping are briefly reviewed here. Due to the short distance ($\sim 2.84$ \AA) NN Fe-Fe bond along the chain direction [Fig.~\ref{structure}(c)], the dominant wave function
overlaps induce hybridization between Fe's $d$ and Se's $p$ orbitals (with Se acting as the Fe-Fe bridge). In this case, the entanglements between orbitals, compatible with a large interorbital hopping $t_{12}$, play a key role in stabilizing the FM order for COFS. This novel conclusion was supported by our second-order perturbation theory analysis and DMRG calculations~\cite{lin2021origin}. Additional results can be found in our previous work~\cite{lin2021origin}.

Next, based on the DMRG results (particularly the spin-spin correlation $S_{i,j}$ and charge fluctuations $\delta{n_{\gamma}}$), we have constructed the phase diagram under hole and electron carrier doping and at $J_H/U = 1/4$, employing COFS as a toy material, and varying $U/W$ and $n$, as shown in Fig.~\ref{phase_n}. Four interesting electronic phases were obtained in our phase diagram: (1) a normal metallic (M) phase, (2) an OSMP1, (3) an OSMP2, and finally at very large $U/W$ (4) a MI state. Note that the boundaries coupling values should be considered only as crude approximations. However, the existence of the four regions shown was clearly established, even if the boundaries are only rough estimations.

\begin{figure*}
\centering
\includegraphics[width=0.8\textwidth]{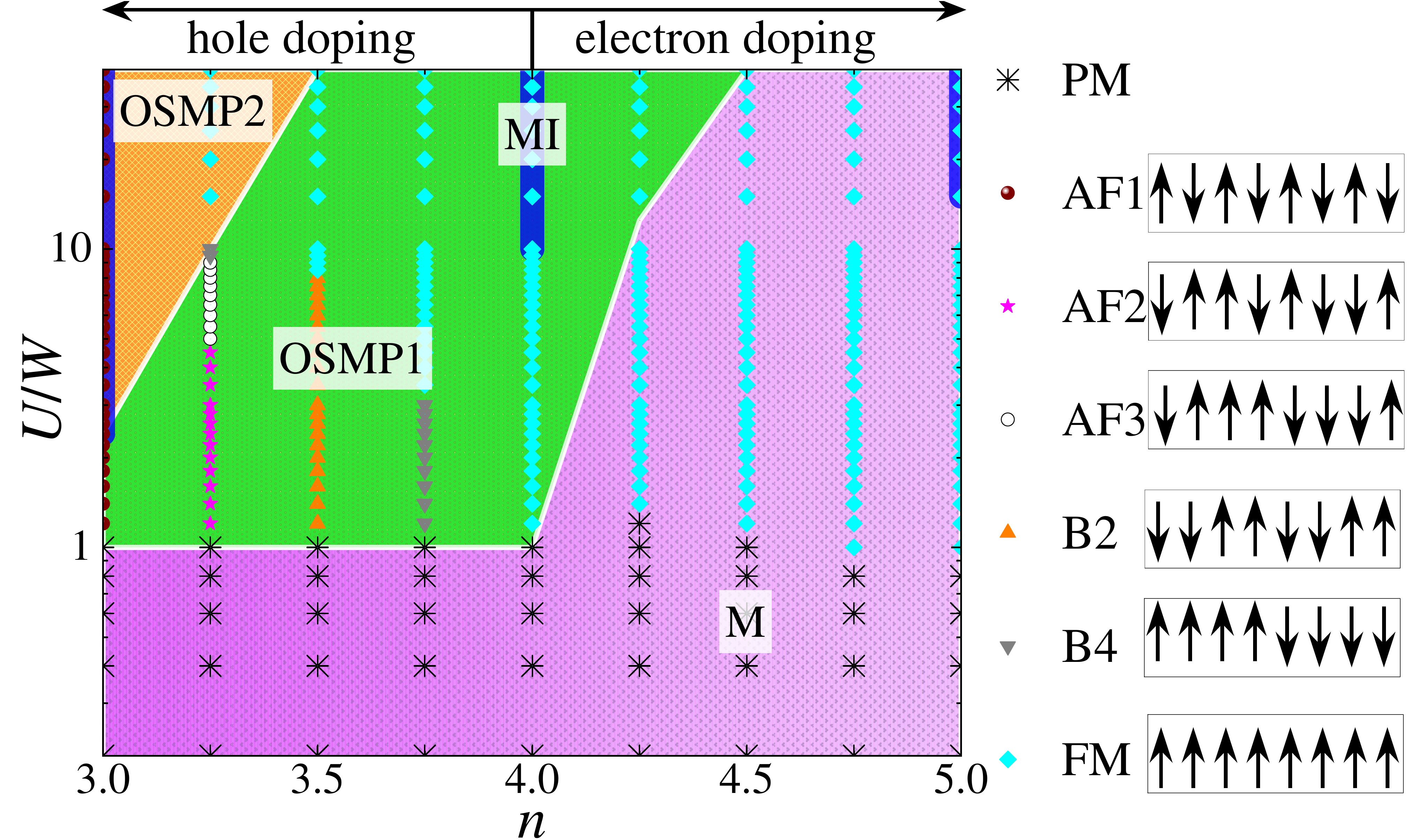}
\caption{Phase diagram of the three-orbital Hubbard model varying $U/W$ and electronic density $n$ (electrons per iron with three orbitals), for an $L=16$ chain with the prototypical value $J_{H}/U = 1/4$.
We define ``hole'' and ``electron'' doping with respect to the $n=4$ prototypical case of many previous studies.
Different electronic phases are indicated by solid colored regions and labels, including metal (M, in pink), orbital selective Mott phase 1 (OSMP1, in green), orbital selective Mott phase 2 (OSMP2, in orange), and Mott insulator (MI in dark blue). These regimes are deduced from the orbital population. Regarding magnetic properties, different magnetic phases are indicated by different symbols and colors for the many points studied  with DMRG, involving paramagnetic (PM), antiferromagnetic 1 (AF1), antiferromagnetic 2 (AF2), antiferromagnetic 3 (AF3), block 2 (B2), block 4 (B4), and ferromagnetic (FM) phases. Block phases represent ferromagnetic clusters that are antiferromagnetically coupled among them. Note that naively AF3 may be interpreted as a B3 state. However, the arrows shown for AF3 are repeated regularly, thus this state contains two consecutive blocks of three sites, followed by a pair of spins up and down. It is not just three-sites blocks.}
\label{phase_n}
\end{figure*}

At small $U/W$ ($\lesssim 1$) and for all electronic densities $n$ investigated, a metallic weakly interacting paramagnetic (PM) state is found, with three itinerant orbitals, where the hopping term plays the leading role for their metallic behavior. At intermediate and strong $U/W$, several magnetic and electronic phases were obtained, depending on the density $n$. At strong $U/W$ and integer $n$, the system is in a Mott insulator because the charges of all orbitals are localized and locked at an integer number either 1 or 2. For $4\lesssim n \lesssim 5$ (electron doping region), all orbitals tend to be itinerant leading to a metallic state indicated by having noninteger $n_{\gamma}$ values. For $3\lesssim n \lesssim 4$ (defined as the ``hole doping'' region taking $n=4$ as the state
of reference) and at intermediate $U/W$, the OSMP1 is obtained with one localized orbital and two itinerant orbitals. For $n$ close to 3 and strong $U/W$, the OSMP2 was observed, with two localized orbitals and one itinerant orbital.

Then, let us discuss the rich magnetic phases unveiled by DMRG in the phase diagram varying the carrier doping. At small $U/W$ ($\lesssim 1$) and all electronic densities $n$, a metallic PM phase is the ground state. For $3\lesssim n \leq 5$ and in the intermediate and strong $U/W$ regions, a fully saturated FM state dominates. At $n = 3$, the canonical staggered AFM phase with the $\uparrow$-$\downarrow$ configuration is observed. Increasing $U/W$ at $n = 3$, the system eventually enters a strongly MI state after the OSMP due to the dominant role of the superexchange Hubbard interaction. For OSMP1 at intermediate $U/W$, four different magnetic configurations were obtained, including AF2, AF3, B2, and B4 magnetic states in the hole-doping region at $n<4$. Those four phases can be simply understood by the competition between FM and AFM tendencies along the Fe-chain that lead to a ``hidden
frustration'' and the development of exotic states. The block phases (B2 and B4)~\cite{rincon2014exotic,rincon2014quantum,herbrych2019novel} were found to be located within the OSMP1 phase at densities $n =3.25$ (B4, $U/W \in [9.5,10]$), $3.5$ (B2, $U/W \in [1.2,8.0]$), and $3.75$ (B4, $U/W \in [1.2,3.0]$).

\subsection{Orbital-selective Mott phases under hole doping $n<4$}

The OSMP state is interesting because it displays simultaneously metallic and insulating bands. In our multi-orbital system, and in the intermediate $U/W$ region, the competition between the non-interacting bandwidth $W$ (corresponding to the kinetic hopping parameter $t$) and electronic correlations (Hubbard $U$, Hund coupling $J_H$) could lead to the orbital-selective localization: one electron with small hopping localizes in
one orbital with growing electronic correlations (Mott-localization) while other orbitals can remain metallic with itinerant bands of mobile electrons because their hoppings are large enough to not become fully localized at intermediate correlations.

\subsubsection{$n=3.25$}

First, let us discuss the OSMP at $n = 3.25$. As shown in Figs.~\ref{n_u1} (a) and (b), at small $U/W$ ($\lesssim 1$), the three orbitals have noninteger $n_{\gamma}$ values with large charge fluctuations (${\delta}n_{\gamma}$), leading to a normal metallic state. The spin-spin correlation indicates a PM state in this region (not shown here), corresponding to an unsaturated spin-squared $\langle {\bf{S}}^2\rangle_{\gamma}$ [Fig.~\ref{n_u1} (c)]. As $U/W$ increases, the ${\gamma} = 0$ orbital population reaches $1$, and that point and beyond there are no charge fluctuations in this orbital, indicating localized electronic characteristics, while the other two orbitals (${\gamma} = 1$ and ${\gamma} = 2$) have noninteger electronic density with some charge fluctuations, leading to metallic electronic features. Furthermore, $\langle {\bf{S}}^2\rangle_0$ saturates at $3/4$, corresponding to a half-filled orbital, while $\langle {\bf{S}}^2\rangle_1$ and $\langle {\bf{S}}^2\rangle_2$ are less than $3/4$. In this case, the system is in the OSMP1 state with one localized orbital and two itinerant orbitals. In the regime of $U/W \textgreater 10$, the occupation number of the ${\gamma} = 1$ orbital also becomes $1$ and the associated
charge fluctuations also vanish ${\delta}n_1 = 0$, as displayed in Figs.~\ref{n_u1} (a) and (b), leading to another insulating band. Then, in this regime, the system has two fully localized orbitals and one itinerant orbital, defining the OSMP2 phase.

\begin{figure}
\centering
\includegraphics[width=0.5\textwidth]{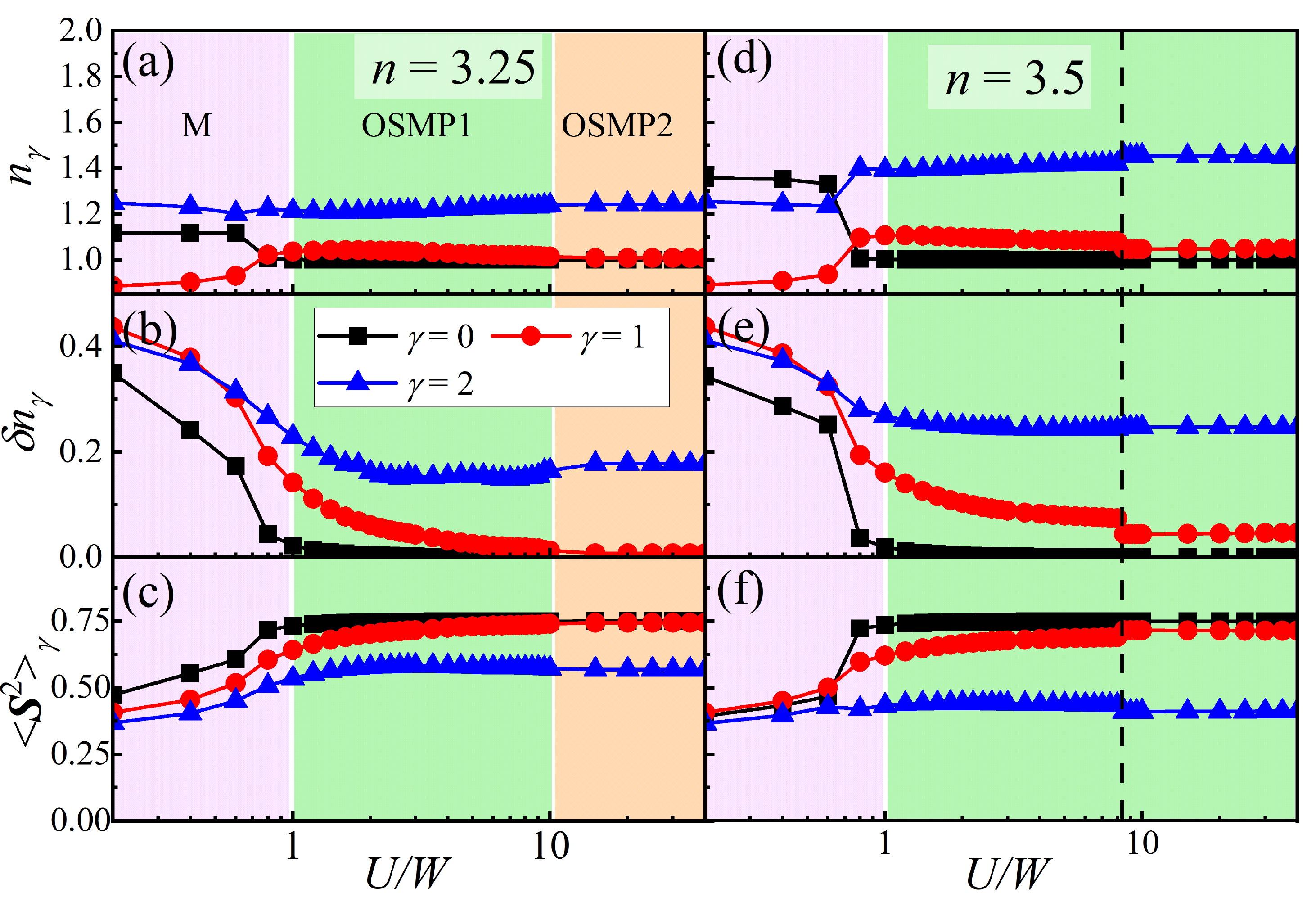}
\caption{Orbital-resolved occupation number $n_{\gamma}$, charge fluctuations ${\delta}n_{\gamma}$, and spin squared $\langle {\bf{S}}^2\rangle_{\gamma}$ varing $U/W$, at fillings $n =3.25$ and $3.5$ and for $J_{H}/U = 1/4$. In the OSMP1 phase at $n=3.5$, the boundary between the block and FM magnetic states is marked by a dashed line.
Note that the precise location of the transitions is difficult to estimate and should be considered only crude
approximations, but the existence of the several different regimes is clear.}
\label{n_u1}
\end{figure}

Those results can be naturally understood by considering different physical parameters. At small $U$, the kinetic portion plays a dominant role, leading to three itinerant metallic orbitals. As $U/W$ increases, the ${\gamma} = 0$ orbital is the first one to be fully localized because the interorbital hopping of orbital ${\gamma} = 0$ is the smallest among the three orbitals. Hence, the system becomes a OSMP1 state.
Because of the large hybridization between ${\gamma} = 1$ and ${\gamma} = 2$, larger than the intraorbital
hopping of ${\gamma} = 0$ itself, robust values of $U/W$ are needed to Mott localize the ${\gamma} = 1$ orbital.
This leads to a large region of OSMP1 in $n = 3.25$. Although ${\gamma} = 1$ and $2$ have comparable hopping terms, the extra $0.25$ electrons favor to occupy the orbital ${\gamma} = 2$ because ${\gamma} = 2$ has a lower crystal-field energy level. Thus, further increasing $U/W$, the ${\gamma} = 1$ orbital becomes Mott-localized as well. Then, as found numerically, the system enters the OSMP2 with two fully localized orbitals and one itinerant orbital, different from OSMP1.

It is interesting to remark that the OSMP2 state was found previously in investigations of the three-orbital Hubbard model although in another context. Specifically, in Ref.~\cite{rincon2014quantum} OSMP2 was reported in a wide range of densities $3<n<4$, while in our case only at $n=3.25$ (at least in the $U/W$ range investigated). The crucial difference between the previous work and the present publication regarding OSMP2 is that the hopping matrix employed before~\cite{rincon2014quantum} tried to mimic the physics of the two-dimensional superconductors translated into a one-dimensional environment, by having hole and electron pockets. In other words, the previous work~\cite{rincon2014quantum} did not use hoppings derived from DFT with a specific material in mind, as in our current
publication with COFS. On one hand, the qualitative similarities suggest that the existence of OSMP2 may be a general feature of this hole doping ($n<4$) portion of the phase diagram when the parent compound is in the OSMP1 state. On the other hand, the differences between the two cases also illustrates that to make concrete predictions about a particular material, such as COFS here, requires using the proper hopping amplitudes. Adding to this conclusion is that in Ref.~\cite{rincon2014quantum} the many exotic magnetic phases to be described below were not reported.

\subsubsection{$n=3.5$ and $n=3.75$}

Next, the DMRG results for density $n = 3.5$ are presented in Figs.~\ref{n_u1} (d-f). Similarly to the results at $n = 3.25$, as $U/W$ increases the ${\gamma} = 0$ orbital quickly reaches a half-filled state with occupation number $n_0 = 1$ without charge fluctuations. Meanwhile, the ${\gamma} = 2$ orbital in the entire range of $U/W$ that we explored has noninteger electronic density with large charge fluctuations, resulting in a strong metallic electronic band. Different from the case of $n = 3.25$, the ${\gamma} = 1$ orbital still keeps small but non-zero charge fluctuations. Hence, we only observed the OSMP1 at density $n = 3.5$, again at least within the coupling range we studied. In addition, we also present the DMRG results for $n = 3.75$ [see Figs.~\ref{n_u2} (a-c)], where only OSMP1 was found in the $U/W$ range studied. By comparing the occupation number and charge fluctuation of the ${\gamma} = 1$ orbital under different electronic densities $n$, we found that the $n_1$ and ${\delta}n_1$ decrease at the same $U/W$ as the electronic density decreases from $n = 3.75$ to $n = 3.25$. Hence, OSMP2 was not observed at $n = 3.5$ and $n = 3.75$ as well, in the range of $U/W$ we studied. We believe that if $U/W$ becomes large enough, the ${\gamma} = 1$ orbital should eventually become localized (thus, OSMP2 should cover a larger portion of parameter space than our phase diagram Fig.~\ref{phase_n} suggests). In addition, at the hole doping ($n<4$) densities reported in this paragraph, several magnetic phases were also obtained. We will discuss those magnetic states in the next section.

\subsubsection{$n>4.0$}

Because the crystal-field splitting of the ${\gamma} = 0$ and ${\gamma} = 1$ are very similar, the electronic doping would affect both orbitals similarly at intermediate $U/W$, leading to two orbitals more than half-filled occupied. Under these circumstances, OMSP should be destroyed under electron doping, namely for density larger than $n=4$. As expected, there is no OSMP already at $n = 4.5$ based on our DMRG calculations. Both ${\gamma} = 0$ and ${\gamma} = 1$ orbitals have more than $1$ electrons with considerable charge fluctuations, leading to two itinerant metallic orbitals, as displayed in Figs.~\ref{n_u2} (d-e).

\begin{figure}
\centering
\includegraphics[width=0.5\textwidth]{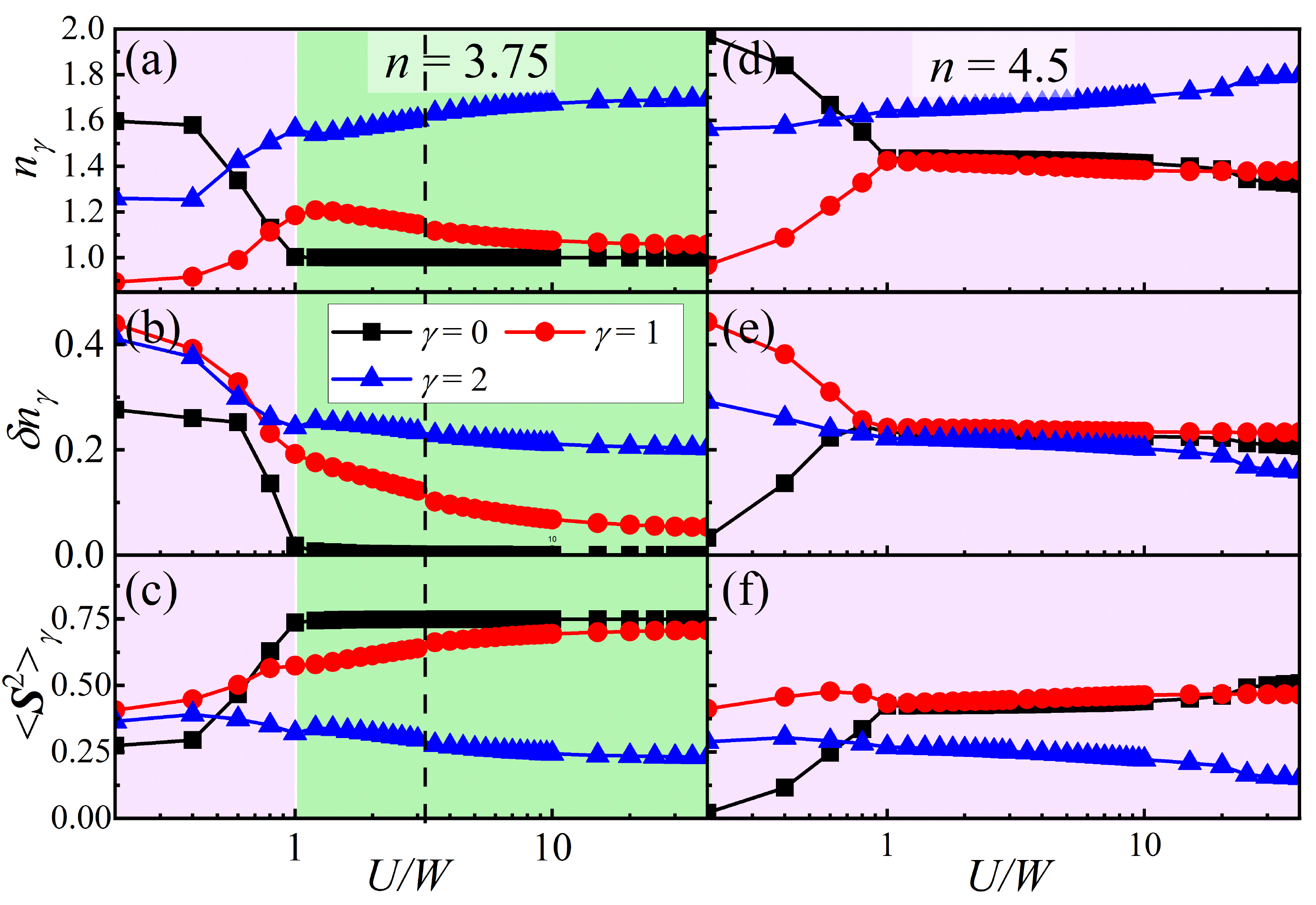}
\caption{Orbital-resolved occupation number $n_{\gamma}$, charge fluctuations ${\delta}n_{\gamma}$, and spin squared $\langle {\bf{S}}^2\rangle_{\gamma}$ varing $U/W$ at fillings $n =3.75$ and $4.5$, and at $J_{H}/U = 1/4$. In the OSMP1 phase for $3.75$, the boundary between the magnetic block and FM states is marked by a dashed line.}
\label{n_u2}
\end{figure}

\subsection{Magnetic phases under hole doping $n<4$}

Finally, let us discuss the many interesting magnetic phases we found at the OSMP1 region under hole doping ($n<4$). Figure~\ref{Sq} (a) shows the spin-spin correlation $S_{i,j}=\langle {\bf{S}}_{i}\cdot {\bf{S}}_{j}\rangle$ vs distance $r$, for different values of $U/W$ at electronic density $n = 3.75$. The distance is defined as $r=\left|{i-j}\right|$, with $i$ and $j$ site indexes. At small Hubbard interaction $U/W \lesssim 1$, the spin-spin correlation $S_{i,j}$ decays rapidly as distance $r$ increases [see result at $U/W = 0.6$ in Fig.~\ref{Sq} (a)], indicating PM behavior. Accordingly, there is no peak appearing for the spin structure factor $S(q)$, as displayed in Fig.~\ref{Sq}(b). As $U/W$ increases to 2.0, the spin-spin correlation $S_{i,j}$ indicates a ${\uparrow}{\uparrow}{\uparrow}{\uparrow}{\downarrow}{\downarrow}{\downarrow}{\downarrow}$ spin configuration, namely the B4 block phase. This  corresponds to the clear peak in $S(q)$ at $q = \pi/4$ [see Fig.~\ref{Sq} (b)]. At larger $U/W$, the system transitions to a FM state, as shown in Figs.~\ref{Sq} (a) and (b) [see results at $U/W = 7.0$ and $10.0$.]

\begin{figure}
\centering
\includegraphics[width=0.5\textwidth]{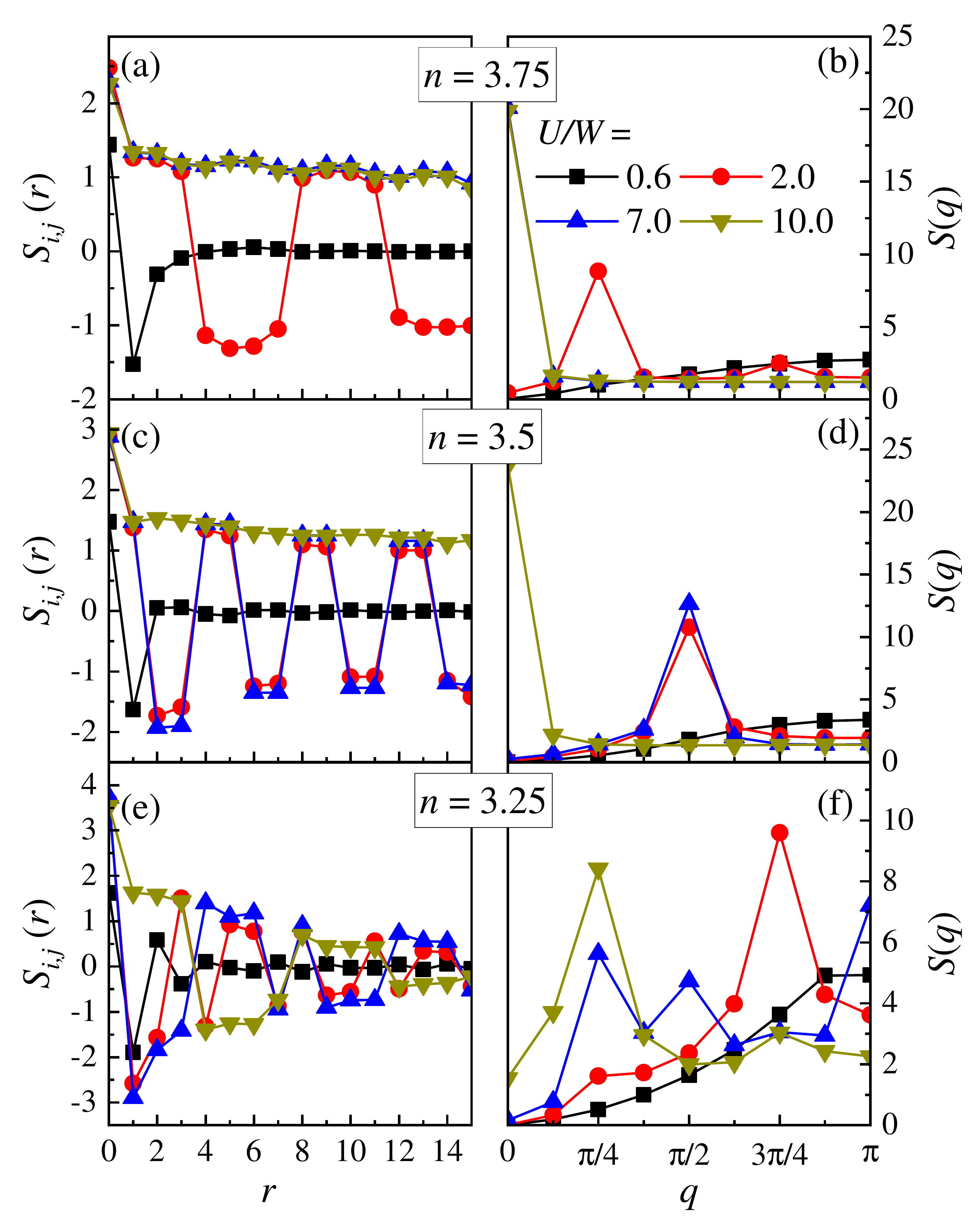}
\caption{The real-space spin-spin correlations and spin structure factor for the indicated values of
$n$ and $U/W$.}
\label{Sq}
\end{figure}

At $n = 3.5$, after the expected small $U/W$ PM state, another interesting block state was observed with
spin configuration ${\uparrow}{\uparrow}{\downarrow}{\downarrow}$ (i.e. the B2 block), as displayed in Fig.~\ref{Sq} (c). The spin structure factor $S(q)$ presents a sharp peak at $q = \pi/2$, corresponding to the latter real-space spin arrangement [see results at $U/W = 2.0$ and $U/W = 7.0$ in Fig.~\ref{Sq} (d)]. By further increasing $U/W$, the system transitions again to the FM phase in the region of our study ($U/W \lesssim 20$).

As hole doping increases, i.e. as $n$ decreases, other magnetic phases were found at $n = 3.25$, such as the AF2 and AF3 states, as displayed in Figs.~\ref{Sq}(e) and (f). However, at large $U/W$, again the system enters the FM state due to strong Hund coupling $J_H$ (not shown).  As discussed before, as
the hole doping transitions from $n = 4$ to $n = 3$, eventually
the $n_2$ electronic population changes from double occupied to single occupied at large $U/W$, thus reducing the tendency toward FM order induced by the $t_{12}$ hopping, and increasing the AFM tendency due to superexchange. In all these cases the many interesting complex magnetic phases can be understood by the competition between AFM and FM tendencies in the intermediate Hubbard coupling regime.

In addition, we also calculated the total charges per site at the three electronic densities $n$ just discussed, as displayed in Fig.~\ref{Nq}. At $n = 3.25$ and $n = 3.75$, we observed charge disproportionation for different magnetic phases as displayed in Figs.~\ref{Nq} (a) and (e). To better understand this disproportionation, we also studied the charge structure factor $N(q)$, shown in the left column of Fig.~\ref{Nq}. At $n = 3.75$, we did not observe any obvious peak in the corresponding charge structure factor, while there is a peak at $\pi/2$ for $n = 3.25$. $N(q)$ also contains the information of charge fluctuations: for example $N(q)$ is suppressed when the $U/W$ increases. In addition, the charge disproportionation pattern in Fig.~\ref{Nq}(a) is not that strong as in Fig.~\ref{Nq}(e), so that the peak is not that obvious in Fig.~\ref{Nq}(b) as in Fig.~\ref{Nq}(f). But in Fig.~\ref{Nq}(b), the peak starts to show up at $\pi/2$ at large $U/W$. At $n = 3.5$, the charge in the various magnetic states remains uniformly distributed.

\begin{figure}
\centering
\includegraphics[width=0.5\textwidth]{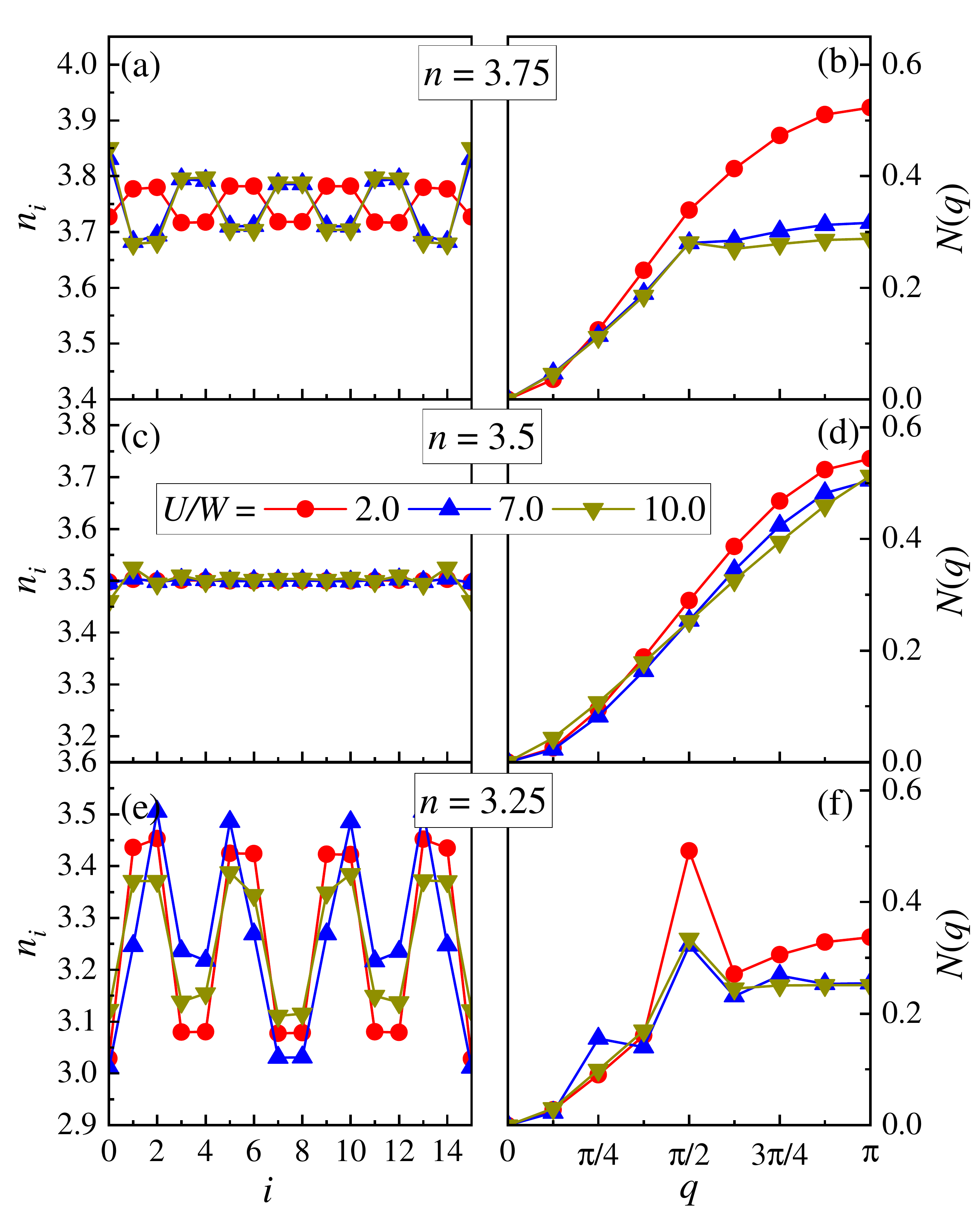}
\caption{Total charge at each site and charge structure factor, for the indicated values of $n$ and $U/W$.}
\label{Nq}
\end{figure}

\section{Conclusion}

In this publication, we have systematically studied the iron chain CeO$_2$FeSe$_2$, with Fe in valence +2 thus with $n = 6$ in the $3d$ orbitals. We have focused on varying the carrier doping away from $n=6$, both via electron and hole doping ($n>4$ and $n<4$, respectively). To make the simulation computationally feasible, we have included three-orbitals in our Hubbard model, instead of five, and have used the DMRG algorithm in order to achieve sufficient accuracy. For the prototypical Hund coupling value $J_{H}/U = 1/4$, we unveiled a rich electronic and magnetic phase diagram varying the electronic density $n$ and electronic Hubbard correlation $U/W$. At small $U/W$ ($\lesssim 1$), for all the range of electronic densities studied, the system is a metallic weakly-interacting paramagnetic state with three itinerant orbitals, i.e. with the hopping kinetic energy dominating the physics. At integer electron number $n$, growing $U/W$ we obtained either a stable FM state for $n =4$ and $5$, or an AFM state at $n = 3$. In the hole-doping region ($n<4$), we observed two kinds of OSMPs: OSMP1 (with one localized orbital and two itinerant orbitals) and OSMP2 (with two localized orbitals and one itinerant orbital). In the electron-doping region above $n=4$, the FM order is dominant with metallic behavior in the entire robust range of $U/W$ that we studied.

We observed that the qualitative difference between the OSMP2 and OSMP1 phases is that in the former the
${\gamma} = 1$ orbital localizes when increasing the electronic correlation $U/W$. Namely,
the observable $n_1$ reaches 1 while ${\delta}n_1$ reaches zero at a critical $U/W$ when $n = 3.25$, leading to the OSMP2 appearing. This does not occur at other densities, in the $U/W$ range investigated. Furthermore, we also found several interesting magnetic phases at $n<4$, involving B2, B4, AF2, and AF3 states (see Fig.~\ref{phase_n}). Furthermore, charge disproportionation was obtained in the B4 phase while the charge of the B2 state remains uniformly distributed.

We believe our theoretical phase diagram will encourage a more detailed experimental study of 1D iron chalcogenide compounds, or related systems, such as COFS and Na$_2$Fe$X_2$. More specifically, here a possible strategy is proposed to modify the chemical formula by ``ion doping'' at the Ce sites, allowing to reach different electronic densities $n$. For example, using Ba$^{2+}$  to replace partially Ce$^{3+}$ at the Ce sites, it would result in a hole doping effect for the system, namely a reduction of the density $n$ away from 4. In more detail, the valence state of Fe in (Ce$_{2-x}$Ba${_x}$)O$_2$FeSe$_2$ will become Fe$^{(2+x)+}$, corresponding to charge density $n=4-x$ in our three-orbital model. For example, $25 \%$ doping with Ba$^{2+}$ randomly replacing Ce will lead to an Fe valence of $+2.5$, corresponding to the line $n = 3.5$ in our Hubbard model phase diagram.  Furthermore, the other limit of electronic doping can also be reached by using Hf$^{4+}$ to replace Ce$^{3+}$. Then, the valence state of Fe in  (Ce$_{2-x}$Hf${_x}$)O$_2$FeSe$_2$ will become Fe$^{(2-x)+}$, which corresponds to $n=4+x$ in our study. For example,  if $25\%$ Hf$^{4+}$ is randomly doped into the Ce sites, the valence of Fe should become $+1.5$, corresponding to the $n = 4.5$ line in our phase diagram Fig.~\ref{phase_n}. Considering the experimental details, we believe hole doping seems to be easier than electron doping. Regardless, crystal growers and neutron scattering experts could confirm our predictions by this ion doping procedure.

\section{Acknowledgments}
The work of L.-F.L., Y.Z., A.M. and E.D. was supported by the U.S. Department of Energy (DOE), Office of Science, Basic Energy Sciences (BES), Materials Sciences and Engineering Division. G.A. was partially supported by the scientific Discovery through Advanced Computing (SciDAC) program funded by U.S. DOE, Office of Science, Advanced Scientific Computing Research and BES, Division of Materials Sciences and Engineering.
J.H. acknowledges grant support by the Polish National Agency for Academic Exchange (NAWA) under Contract No. PPN/PPO/2018/1/00035 and by the National Science Centre (NCN), Poland, via Project No. 2019/35/B/ST3/01207.

\bibliographystyle{apsrev4-1}
\bibliography{ref3}
\end{document}